\documentstyle[epsf,rotate]{mn}

\newcommand{\eg}{{\sl e.g. }}
\newcommand{\ie}{{\sl i.e. }}
\newcommand{\etal}{{\sl et al. }}
\newcommand{\Msun}{\mbox{$M_{\odot}$}}

\title[An ``outside-in'' outburst of Aql X--1]
{An ``outside-in'' outburst of Aql X--1}
\author[Shahbaz et al.]{
T. Shahbaz,$^{1}$ R.M. Bandyopadhyay,$^{1}$ P.A. Charles,$^{1}$ R.M. Wagner,$^{2}$ P. Muhli,$^{3}$
\and P. Hakala,$^{3}$ J. Casares,$^{4}$ and J. Greenhill$^{5}$\\
$^{1}$University of Oxford, Department of Astrophysics, Nuclear Physics Building, Keble Road, Oxford, OX1 3RH, UK \\
$^{2}$Department of Astronomy, Ohio State University, 174 West 18th Ave., Columbus, OH 43210-1106, USA \\
$^{3}$Observatory, P.O. Box 14, FIN-00014 University of Helsinki, Finland \\
$^{4}$Instituto de Astrof\'\i{}sica de Canarias, 38200 La Laguna, Tenerife, Spain\\
$^{5}$Physics Department, University of Tasmania, GPO Box 252C, Hobart, Tasmania, Australia
}

\begin{document}

\maketitle

\begin{abstract}

\noindent
We present optical spectroscopy and optical and infrared photometry of
the neutron star soft X-ray transient Aql X--1 during its X-ray outburst
of August 1997.  By modelling the X-ray, optical, and IR light curves, we
find a 3 day delay between the IR and X-ray rise times, analogous to the
UV-optical delay seen in dwarf novae outbursts and black hole X-ray
transients. We interpret this delay as the signature of an ``outside-in''
outburst, in which a thermal instability in the outer disc propagates
inward.  This outburst is the first of this type definitively identified
in a neutron star X-ray transient.

\end{abstract}


\section{Introduction}

Aquila X--1(=V1333 Aquilae) is known to undergo regular X-ray and optical
outbursts on a timescale of $\sim$1 year (Kaluzienski et al. 1977;
Priedhorsky \& Terrell 1984; Charles \etal 1980), much more frequently
than the other neutron star transient Cen X--4 (McClintock \& Remillard
1990). Aql X--1 therefore presents us with regular opportunities to test
the models proposed to explain the X-ray outbursts occurring in these
low-mass X-ray binary transient systems.

Aql X--1 also exhibits type 1 X-ray bursts (Koyama 1981; Czerny, Czerny
\& Grindlay 1987), indicating that the compact object is a neutron star.
Observations in quiescence have shown that the mass-donating companion is
a V=19.2 K1 $\sc IV$ star (Shahbaz, Casares \& Charles 1997).  The
optical counterpart brightens by $\sim$2--5 magnitudes during X-ray
outbursts, interpreted as reprocessing of radiation in the accretion disc
(Thorstensen, Charles \& Bowyer 1978; Canizares, McClintock \& Grindlay
1980; Charles \etal 1980; van Paradijs \etal 1980).

The RXTE All Sky Monitor recorded an X-ray outburst of Aql X--1 between
late January and early March 1997 (Levine \& Thomas 1997).  By 30 March
1997 Aql X--1 was reported to be optically in quiescence (Ilovaisky
\& Chevalier 1997).  Aql X--1 then had another outburst in August 1997
(Charles \etal (1997) reported that it had brightened to V=17.8 on August
7 1997), reaching a maximum level at around V=17.5 (Chevalier \&
Ilovaisky 1997).  In this letter we report on X-ray/optical/infrared
photometry of Aql X--1 obtained during the August 1997 outburst.

\section{Observations and Data Reduction}

A complete journal of the infrared and optical photometric observations is 
presented in Table 1.

\vspace{-4mm}

\subsection{Infrared Photometry}

\subsubsection{Lowell Observatory}

We obtained $K$-band images of Aql X--1 over a total of 16 nights in 1997
July and August using the Ohio State Infrared Imager/Spectrometer (DePoy
\etal 1993) on the Perkins 1.8-m telescope of the Ohio State and Ohio
Wesleyan Universities at Lowell Observatory.  The 210 images were taken
with the $f/7$ camera which provides a 2\farcm7 field of view at a
resolution of 0\farcs63 pixel$^{-1}$; the seeing was typically 2\farcs9.

A standard observing sequence consisted of five consecutive images of 60
seconds for Aql X--1; the position of the object on the array was
moved between exposures, so that the group could be median-stacked to
produce a sky frame.  The images were processed by first running a simple
interpolation program to remove bad pixels.  A median-combined sky image
was then created from the five exposures and subtracted from each image.
The frames were then flat-fielded using images of a tungsten lamp on a
dome flat.  These were taken with the lamp on and also with the lamp off
so that the thermal component could be subtracted from the flat-field.

In order to determine the counts for stars in each image we first
mosaiced each group of five images so that we could obtain more accurate
photometry.  The counts for Aql X--1 and a local standard were then
computed using the aperture photometry routine described in Shahbaz,
Naylor \& Charles (1994) with a 3 pixel radius aperture.  The magnitudes
of Aql X--1 are given relative to a bright star 25 arcsec West and 4
arcsec South of Aql X-1 (Shahbaz \etal 1998). Typical errors in $K$ were
0.08 mags.

\subsubsection{UKIRT}

We obtained $JHK$-band images of Aql X--1 on the 16th and 18th July 1997
using the InSb infrared array IRCAM-3, on the 3.8-m United Kingdom
Infrared Telescope atop Mauna Kea, Hawaii. Photometric conditions also
allowed the acquisition of UKIRT standard stars. A typical observing
sequence consisted of eight consequtive images of 10s for Aql X--1, where
the object's position on the array was moved between exposures, so that
the group could be median stacked to produce a flat-field frame. The
standard reduction procedures were then followed.

The counts for Aql X--1 and a local standard were then computed using the
aperture photometry routine described in Shahbaz, Naylor \& Charles
(1994). The magnitude of the local standard was measured, which then
allowed us to determine the apparent $JHK$ magnitude for Aql X--1 in
quiescence. We obtained $J$=16.59$\pm$0.02, $H$=16.05$\pm$0.03 and
$K$=15.91$\pm$0.04.

We could not determine the colour correction between the UKIRT and Lowell
filter systems. Therefore we do not use the UKIRT $K$-band data 
in the analysis of the IR light curve of Aql X--1.

\subsection{Optical Photometry}

\subsubsection{Teide Observatory}

We obtained photometry on the 80cm telescope IAC80 on the nights of
1997 August 12-16, 20-21 and 30 and 1997 September 11 and 15-16.
The frames were acquired with a Thomson chip which has a 7.5x7.5
arcmin$^2$ field of view and a scale of 0.43 arcsec pixel$^{-1}$.  Johnson
$B$ and $V$ images were obtained using exposure times of 900s and 600s 
respectively; the seeing was 0.8-1.5 arcsec.
Typical errors were 0.10 and 0.05 mags in $B$ and $V$ respectively.

\vspace{-2mm}

\begin{table}
\small{
\caption{Log of Optical/IR Photometry}
\begin{center}
\begin{tabular}{lllc}
Location  & Month  	& Day                           & Bands   \\ 
		&		&		&	\\

Lowell 	    & July   	& 16,17,18,19,20,23,24,25,26,31 & $K$ \\
            & August 	& 1,12,13,14,20,23              & $K$ \\
UKIRT       & July   	& 16, 18                        & $K$ \\
            &        	&              	       &     \\ 
La Palma    & July   	& 30      	       & $V$ \\
	    & August 	& 5,6,7,8 	       & $U, B, V$ \\
	    & August 	& 5,6     	       & $R, I$ \\
            &        	&         	       &     \\
Tasmania    & August 	& 19, 23  	       & $V, I$  \\
            &           &         	       &     \\
Teide       & August    & 13,14,15,16,21,22,30 & $B, V$ \\
	    & September & 11,15,16             & $B, V$ \\
	    & October   & 9,10                 & $B, V$ \\ 
\end{tabular}
\end{center}
}
\end{table}

\subsubsection{La Palma}

On 1997 August 5-8, $UBV$ images were taken at the 2.6-m Nordic Optical
Telescope (NOT) at the Observatorio del Roque de los Muchachos on La
Palma; one image was taken in each filter on each of the four nights.  We
also have one $V$ observation from July 30 as well as $R$ and $I$
observations from August 5 and 6.  We used a thinned, coated Loral-Lesser
2000x2000 array with typical exposure times of 120 sec in $U$, 60 sec in
$B$ and $30$ sec in $V$. 
Typical errors were 0.01 and 0.05 mags in $B$ and $V$ respectively.

\vspace{-2mm}

\subsubsection{Tasmania}

The observations were made on 1997 August 19 and 23 using the Canopus 1-m
telescope of the University of Tasmania at a focal ratio of f/11.  A ST6 CCD 
photometer with Cousins $V$ and $I$ filters was used with exposures of 5 and 
10 minutes.  

\vspace{4mm}

\noindent
All of the optical images have undergone standard CCD reduction, including 
bias subtraction, flat-fielding, and removal of bad pixels.  Aperture 
photometry was performed on Aql X-1 and a local standard using the  $\sc ark$ 
`APPHOT' routine; the apertures used ranged from 3-5 pixels in radius, 
varying with the different instruments and seeing conditions.  Magnitudes of 
Aql X-1 are given relative to the same bright star used for the IR photometry.

\subsection{Optical Spectroscopy}

Aql X--1 was observed with the ISIS spectrograph on the 4.2-m
William Herschel telescope on the nights of 1997 August 6-7.  Two 1000s
spectra were obtained covering $\lambda\lambda$4110-4910 
and $\lambda\lambda$5900-6700 for the blue and red arms respectively.  A
slit width of 0.8 arcsec was used, which when combined with the 600
line/mm grating resulted in a spectral resolution of 1.1 \AA (FWHM=50
km~s$^{-1}$ at H$\alpha$) for the red arm and 1.1 \AA (FWHM=68
km~s$^{-1}$ at H$\beta$) for the blue arm.

The bias level was removed from each image using the mean overscan
regions and then flat-fielded by using a tungsten lamp.
One-dimensional spectra were extracted using the standard optimal
extraction routines in the PAMELA reduction package, which weights the
pixels along the spatial profile in order to obtain the maximum
signal-to-noise ratio (Horne 1986); wavelength calibration was performed 
using the MOLLY package.

\subsection{$RXTE$ Observations}

The $RXTE$ All-Sky Monitor (ASM) has been operating more or less
continuously since 1996 February 21, providing roughly five to ten scans
of a given source per day in the 2-12 keV energy range.  We obtained the
one-day average X-ray data for Aql X--1 from the public archive
maintained by the $RXTE$ Guest Observers Facility.  For further details
about the instrument and the methods used in the ASM data reduction and
error calculations, see Levine \etal (1996).

\begin{table}
\caption{$URI$ Optical Photometry.
Typical errors are 0.04, 0.03 and 0.01 mags
for $U$, $R$ and $I$ respectively.}
\begin{center}
\begin{tabular}{lcclcc}
Band  & Date$^{a}$ & $\Delta$mag & Band  & Date$^{a}$ & $\Delta$mag\\
      &            &             \\
U     &  666.4490  & -1.23   & I     &  666.4564  &  1.74  \\
      &  667.3827  & -1.42   &     &  667.3865  &  1.76  \\
      &  668.3835  & -1.40   &     &  679.5722  &  1.48  \\
      &  669.5969  & -1.77   &     &  679.5770  &  1.51  \\
      &            &         &     &  683.4890  &  1.36  \\
R     &  666.4548  &  1.40   &     &  683.4957  &  1.29  \\ 
      &  667.3858  &  1.44  & & & \\
      &            &        & & & \\
\end{tabular}
\end{center}
\noindent $^{a}$ HJD - 2,450,000.\\
\end{table}

\section{Results}

\subsection{The distance}

The distance to Aql X--1 can be derived using the apparent $K$-band
magnitude and the surface brightnes $S_{K}$ of the companion star (Bailey
1981). Using $V$=19.2 (Thorstensen et al. 1978) and allowing for an
accretion disc contamination in the range 0--50 per cent and reddening
of $E_{B-V}$=0.35 mags (Shahbaz et al. 1996), we obtain $V_{o}$ in the
range 18.1--18.9.

Using our UKIRT $K$-band magnitude of $K$=15.9 (section 2.1.2) and
assuming no disc contamination in the IR (the disc contamination at
6600\AA\ is only 6 per cent: Shahbaz, Casares \& Charles 1997) we find
$K_{o}$=15.8. Given the limits for the intrinsic colour of the secondary
star and its surface brigthnes, $(V-K)_{o}$=2.30--3.05 and
$S_{K}$=3.58--3.83 respectively (Ramseyer 1994), and using a secondary
star mass of 0.15 \Msun (by comparison with Cen X--4; Shahbaz, Naylor \&
Charles 1997), we find distance values of 2.2--2.4 kpc (note that the
range quoted is due to the uncertainty in the disc contamination in the
$V$-band). We conclude that the distance to Aql X--1 is 2.3$\pm$0.1 kpc,
which is consistent with the previous distance estimate of 2.5 kpc
(Charles et al. 1980).

\subsection{Optical spectrum}

Figure 1 shows the optical outburst spectrum of Aql X--1, which exhibits 
emission features of H$\alpha$ (EW=5.0$\pm$0.5\AA), 
H$\beta$ (EW=2.8$\pm$0.5\AA), H$\gamma$ (EW=4.2$\pm$0.5\AA), 
He$\sc ii$ 4686\AA (EW=3.3$\pm$0.4\AA) and the Bowen blend 4640-4650\AA 
(EW=2.0$\pm$0.3\AA).  There is
also some evidence for an outflow in the system, presumably arising from
an accretion disc wind, as the H$\beta$ emission line has a P-Cygni type
profile, where the blue side of the line profile is absorbed.  Finally,
the emission lines are single-peaked, suggesting that the binary
inclination is low.

\subsection{X-ray light curve}

The August X-ray outburst light curve of Aql X--1 has three distinct
sections (see Fig 2). Initially, the X-rays are constant while the source
is quiescent.  The X-rays then rise linearly to maximum; the subsequent
decay is also linear.  The three boundaries on the curve are the time of
the initial rise, the time of maximum, and the end of the decay.  Using
these three boundaries, we simultaneously fitted the data with a three
component model; the resultant parameters are given in Table 2.  The
secondary maximum feature present $\sim$22 days after the outburst is also
seen in other SXTs (Chen, Shrader, \& Livio 1997) and was removed from the 
fitting procedure.  A detailed discussion of the secondary maxima will be 
presented in Shahbaz, Charles \& King (1998).

\subsection{Optical/IR light curve}

The $V$- and $B$-band light curves of Aql X--1 display three distinct
phases (see Fig 2). At first, the optical flux rises linearly.  However,
instead of peaking at X-ray maximum, the $V$ and $B$ light curves flatten
when the X-rays have reached slightly less than half of the maximum
intensity.  The optical light curve remains in this plateau state until
just after the secondary maximum in the X-ray decay (Shahbaz, Charles \&
King 1998), at which point both
$V$ and $B$ begin a linear decay.  By simultaneously fitting the optical
light curves with the three-part X-ray model described above, we
determine the slopes of the linear rise and decay, and also the start and
end of the plateau region.  The four $U$-band observations also show a
linear rise similar to that seen in the $B$ and $V$ light curves. $R$ and
$I$ measurements were also obtained; the $URI$ relative magnitudes are
given in Table 2.

After an initial period of quiescent observations, the $K$-band light
curve shows a linear rise and then a decrease in flux which may or may
not be associated with the decay.  We only fit the initial rise of the
light curve and determine the slope and start of the rise.  It is
interesting to note that the rate of the increase in brightness is
largest in $V$ and smallest in $K$.  However, since the time coverage is
sparse, there is some uncertainty in this conclusion. The fit parameters
determined for the $V$, $B$, $K$, and X-ray light curves are listed in
Table 2.

Our fits to the light curves show that the time of initial rise is earliest 
in $K$ and latest in X-rays.  The beginning of the rise in the X-ray intensity 
occurred on UT July 31 14.4, 3 days after the initial rise in the $K$-band 
(UT July 28 12.0).  This time-delay is significant at the 98.69 per cent level.

In the middle panel of Fig. 2 we show the hardness ratio 
(3.0--12.0keV/1.3--3.0keV) of the ASM data, which resembles the ``flat topped''
optical light curves ($B$ and $V$).  The unusual shape of the optical light
curves, which level off rather than following the soft X-ray light curve to
its peak, suggest that the flux may arise instead from reprocessing of the
hard X-rays. However, inspection of BATSE, ASM, and optical 
data of two previous outbursts (1996 June and 1997 February) casts doubt on 
this hypothesis.  During the 1996 June event, BATSE recorded a large increase
in hard X-rays, while the ASM light curve showed only a small, erratic
increase in soft X-rays.  In contrast, the 1997 February event was strong
in soft X-rays but was not detected by BATSE.  The optical light curves
of Aql X--1 from both events, however, were very similar, and neither 
showed the ``flat-topped'' behaviour seen in our $V$- and $B$-band data
(Garcia 1988).  It is therefore unclear why our optical light curves do not
appear to show direct reprocessing of the entire X-ray flux.
Spectral information during the entire outburst would help in determining
the origin of the optical flux.

\begin{table}
\begin{center}
\caption{Fitted parameters for the August outburst light curves of Aql
X--1.}
\begin{tabular}{llr}
          &                              &       \\ 
$K$-band  &  Slope of linear rise (mags day$^{-1}$) &  -0.07 $\pm$ 0.05   \\
          &  Start of initial rise$^{a}$   & 658.0 $\pm$ 1.1 \\
          &                                  &       \\ 
$V$-band  & Slope of linear rise  (mags day$^{-1}$) &  -0.14 $\pm$ 0.01   \\
  	  & Slope of linear decay (mags day$^{-1}$) &   0.07 $\pm$ 0.01   \\
	  & Start of plateau$^{a}$   	    & 669.6  $\pm$ 0.5\\
	  & End of plateau$^{a}$   	    & 700.0  $\pm$ 1.1\\
          &                                  	    &       \\ 
$B$-band  & Slope of linear rise  (mags day$^{-1}$) & -0.11$\pm^{+0.12}_{-0.05}$ \\
  	  & Slope of linear decay (mags day$^{-1}$) & 0.09 $\pm$ 0.01   \\
	  & Start of plateau$^{a}$   	    & 669.6  $\pm$ 0.5\\
	  & End of plateau$^{a}$   	    & 701.9  $\pm$ 1.6\\
	&                                                   &       \\
X-rays	& Slope of linear rise (cts s$^{-1}$day$^{-1}$)  &  0.76 $\pm$ 0.03\\
       	& Start of initial rise   & 661.0 $\pm$ 0.5    \\
	& Time at maximum light  & 683.7 $\pm$ 0.3    \\
	& Slope of linear decay (cts s$^{-1}$day$^{-1}$) & -0.51 $\pm$ 0.01 \\
\end{tabular}
\end{center}
\noindent $^{a}$ HJD - 2,450,000.\\
\end{table}

\section{The Observed Delay}

The X-ray delay observed in the outburst of Aql X--1 is analogous to the
well known UV delay observed for dwarf novae (Warner 1995, and references
therein). For dwarf novae the rise in the UV flux starts about 5-15
hours after the beginning of the optical outburst. In the framework of the
standard disc instability model (DIM) one can interpret the UV delay as due to 
an ``outside-in'' outburst.  According to the DIM (\eg Cannizzo 1993, and
references therein), a thermal instability in the outer disc creates a heating 
front which propagates inward.  This front transforms the disc from a cold
(quiescent) state to a hot state.  Because the UV flux is mainly emitted
close to the white dwarf, one expects a delay in its rise equal
to the time it takes the front to travel from the outer disc to the
white dwarf.  In the DIM, however, the calculated travel time of the
front is much shorter than the observed UV delay time ($\sim$ 1 day;
Pringle, Verbunt \& Wade 1986; Cannizzo \& Kenyon 1987). Thus, in this
standard form the model fails to explain the UV delay.

Two solutions have been proposed in order to rescue the DIM, both of
which invoke a central hole in the accretion disc.  An inward moving heating 
front would have to stop at the edge of such a hole; the hole would
then fill up on a viscous timescale, which is much longer than the heat
front propagation time.  Livio \& Pringle (1992) suggested a mechanism
for creating such a hole involving the magnetic field of a weakly magnitized
white dwarf, which can disrupt the inner accretion disc.  They showed that 
such a model can reproduce the UV-delay observed in dwarf novae outbursts.
Meyer \& Meyer-Hofmeister (1994) proposed a different scheme for quiescent
accretion onto a white dwarf that also results in a central hole.  They
invoke inefficient cooling in the disc's upper layers, leading to the
formation of a hot corona and ultimately to the evaporation of the inner
disc. 

Whether a hole is created by magnetic fields or by evaporation, the
effect on the outburst of a dwarf novae is similar.  When the heating front
arrives at the inner edge of the truncated disc it cannot propagate any
further; the (surface) density contrast slowly fills up the hole on a
viscous timescale, thereby producing the required delay of the UV
outburst.

Our photometry suggests that the instability that formed the August 1997
outburst started in the outer regions of the disc and propagated inwards
(an ``outside-in'' event), since the brightening started in the $K$-band
first and then later in X-rays.  The strength of the magnetic field of
the neutron star in Aql X--1 has never been determined from observations,
although it is believed to be weak, as inferred from the lack of any
detectable pulsed X-ray flux in its persistent emission.  It is easily
conceivable that a hole could be created in the accretion disc, either by
the weak magnetic field of the neutron star or by quiescent X-rays
($L_{x}\sim4\times 10^{32}$ ergs~s$^{-1}$; see Verbunt et al. 1994)
evaporating the inner regions of the disc.  Therefore, as in the case for
dwarf novae, one can reproduce the observed time delay between the start
of the optical and X-ray outburst.  The fact that Aql X--1 has a binary
separation 25 percent larger than typical dwarf novae may explain why the
IR--X-ray lag in Aql X--1 is longer than the optical-UV lag in the dwarf
novae.  Detailed computations would be required to see if the 3-day delay
between the IR and X-rays can be explained by the DIM model.

Recently Narayan, McClintock \& Yi (1996) have proposed a model for the
quiescent state of the black hole transients.  In the inner region of
this model the flow is advection-dominated, \ie it is hot and optically
thin and most of the thermal energy is transported across the event
horizon, thereby greatly reducing the observed luminosity for the given
accretion rate. This is because at these temperatures the radiative
cooling timescale is significantly lower than the infall timescale (Narayan et
al 1996). For a neutron star system such as Aql X--1 this energy would
eventually be radiated from the neutron star surface. The difference
between the black hole and neutron star cases has been cited by Narayan,
Garcia \& McClintock (1997) as evidence for the nature of the compact
object.

The key point for Aql X--1 is that the advection-dominated accretion flow
(ADAF) region would in effect act as a ``hole'' in the middle of the
disc, since the standard thin disc would be truncated at the edge of the
advection-dominated flow.  The inner disc becomes advection-dominated at
the end of the {\it previous} outburst when the accretion rate drops
below a certain level. We can then estimate the delay we would expect to
see by scaling the delay observed in GRO~J1655-40 to that in Aql X--1.
The $\sim 6$~day delay seen in GRO~J1655-40 can be modelled as the
diffusion of the inner edge of the disc to the black hole, starting at
$R_{in}=2.4 \times 10^{10}$ cm (Hameury \etal 1997).  The FWZI of the
H$\alpha$ emission profile is $\sim$1500~km~s$^{-1}$, so for Aql X--1,
$R_{in}=1.4\times 10^{10}$cm. If the diffusion rate is the same, we would
expect a delay of $\sim 3$~days (Garcia 1998); this compares well with
the observed time delay between our IR and X-ray light curves.

\section{The propeller effect in Aql X--1?}

We now address the observed X-ray properties of Aql X--1, one of only
two neutron star SXTs (the other is Cen X--4) and with a very short
outburst recurrence time (typically a year or less).  In 1997 outbursts were 
observed in Aql X--1 in Feb. and Aug., the latter of which has been discussed
herein.  The former outburst was the subject of a more detailed XTE study by
Zhang, Yu \& Zhang (1998).  In particular, Zhang \etal found that the X-ray
spectral behaviour (as evidenced by its hardness ratio) of Aql X--1
was constant through the decline until just before it entered quiescence,
at which point the spectrum hardened dramatically (see Fig. 3).

Zhang \etal interpreted this spectral change in terms of the ``propeller'' 
effect.  This effect has been proposed by Stella, White \& Rosner (1986) to 
account for the cessation of X-ray emission from rapidly spinning pulsars in 
Be systems when the mass transfer drops below a certain threshold.  This is 
due to matter entering the rapidly spinning magnetosphere, at which radius 
the rotation speed exceeds the keplerian speed and hence the matter is
ejected rather than accreted.  Zhang \etal propose that Aql X--1 is a spun-up 
pulsar with a weak magnetic field which exhibits the propeller effect when the
accretion rate is sufficiently low.
Note that kHz QPOs have now been clearly seen by Cui et al. (1998) during
the rising phase of the February 1998 outburst.

If the X-ray light curves and spectral properties of other SXTs are
examined carefully (e.g. Chen, Shrader, \& Livio 1997), then very similar
spectral changes at the same stage are seen in A0620-00, GS~1124-683 and
GS~2000+25, all of which are (dynamically) established black hole systems
in which this mechanism could {\bf not} take place.  We also note that
this spectral change occurs somewhat after the secondary maximum in the
light curve.  This is true of the other systems as well, but the
observations are too sparse to demonstrate any systematic effect in this
delay.  There is some evidence for a secondary maximum in the ASM/PCA
light curves at $\sim$18 days after the outburst peak.  In Fig. 3 we show
the X-ray light curves of GS~1124-683 and Aql X--1.

In addition to the propeller effect, we propose that other properties of the
SXTs can be called upon to account for the spectral hardening.  It occurs
late in the outburst, after the secondary maximum, the point at which the
outer irradiated regions of the disc have been accreted and so, barring
the provision of substantial additional material from the secondary, the
rate of accretion will be declining.  This will provide an opportunity
for the remaining inner disc material to be evaporated by the
X-radiation, thereby producing the inner hot, low density region which
can lead to an ADAF and hence a spherical flow onto the neutron star with
a dramatic change in spectrum (but note that being a neutron star system,
the accreted luminosity is of course then radiated as it hits the
surface).

\section{Conclusions}

We have determined a 3 day delay between the IR and X-ray rise times,
analogous to the UV-optical delay seen in dwarf novae outbursts and black
hole X-ray transients. We interpret this delay as an ``outside-in''
outburst, in which a thermal instability in the outer disc propagates
inward. We suggest that an ADAF region was created at the end of the
previous outburst. The evidence for this comes from the observed
hardening of the X-ray spectrum as the system declined into quiescence, a
hardening which is also seen in the black hole soft X-ray transients.
This ADAF region would appear as a ``hole'' in the inner accretion disc,
thereby causing the X-ray-IR delay when the system starts its next
outburst; the hole is needed in order to explain the observed X-ray -- IR
delay.

\section*{Acknowledgments}

We would like to thank Andy Stephens for assisting with the early stages
of the IR data analysis, and Ray Bertram for performing some of the IR
observations. We also thank Miquel Serra for supporting the service
observations of the IAC80 and Tim Naylor for obtaining the UKIRT images.
JC acknowledges support by the Spanish Ministry of Science Grant PB
1995-1132-02-01. The data reduction was carried out on the Oxford
Starlink node using the $\sc ark$ software.

\begin{figure*} 
\rotate[l]{\epsfxsize=500pt \epsfbox[00 00 700 750]{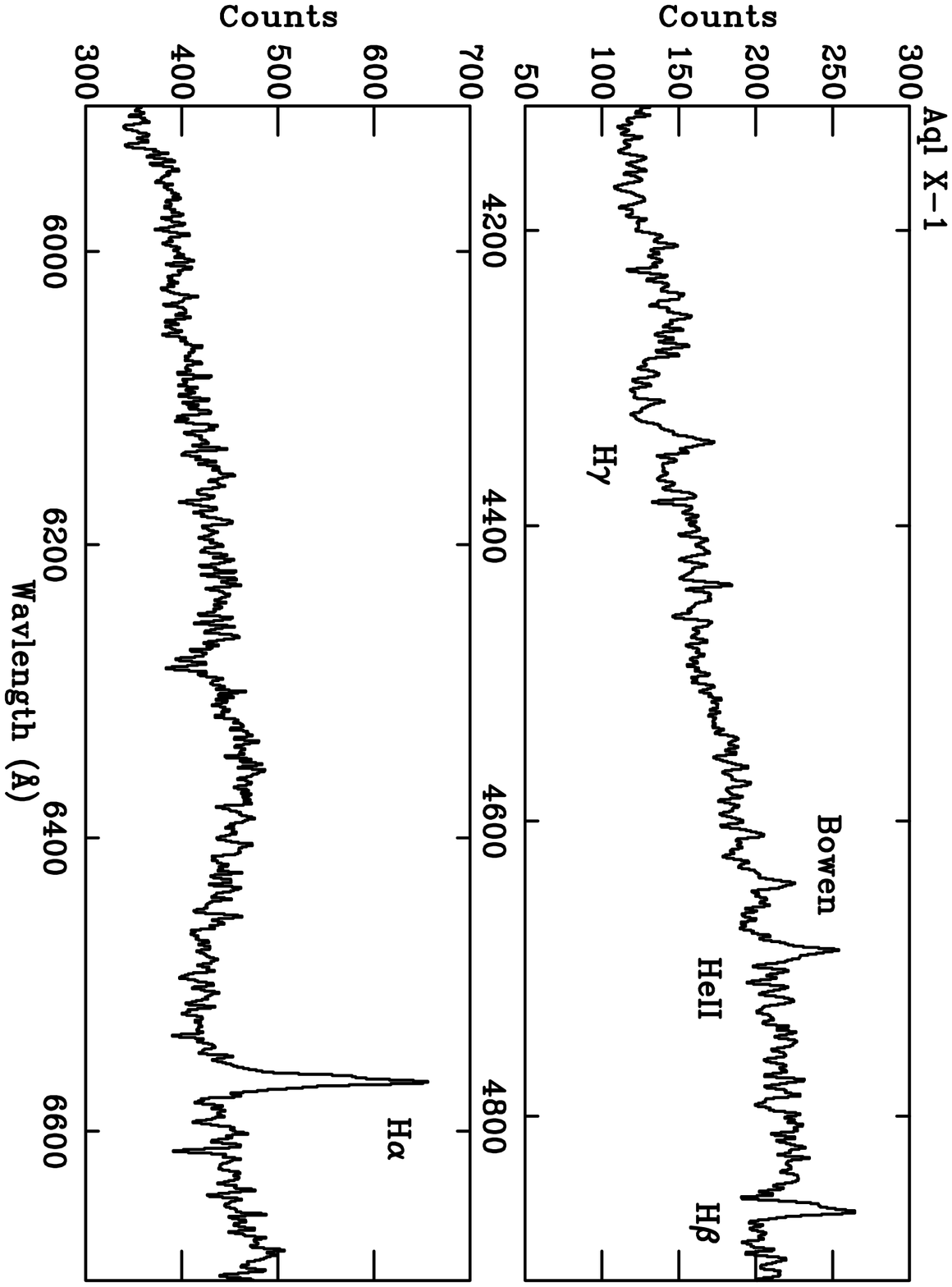}}
\caption{
The optical outburst spectrum of Aql X--1. 
The Balmer (H$\alpha$, H$\beta$ and H$\gamma$) and He$\sc ii$ (4686\AA)
emission features are present and also the Bowen blend
4640-4650\AA. The H$\beta$ emission line has a P-Cygni type profile.}
\end{figure*}

\begin{figure*} 
\rotate[l]{\epsfxsize=500pt \epsfbox[00 00 700 750]{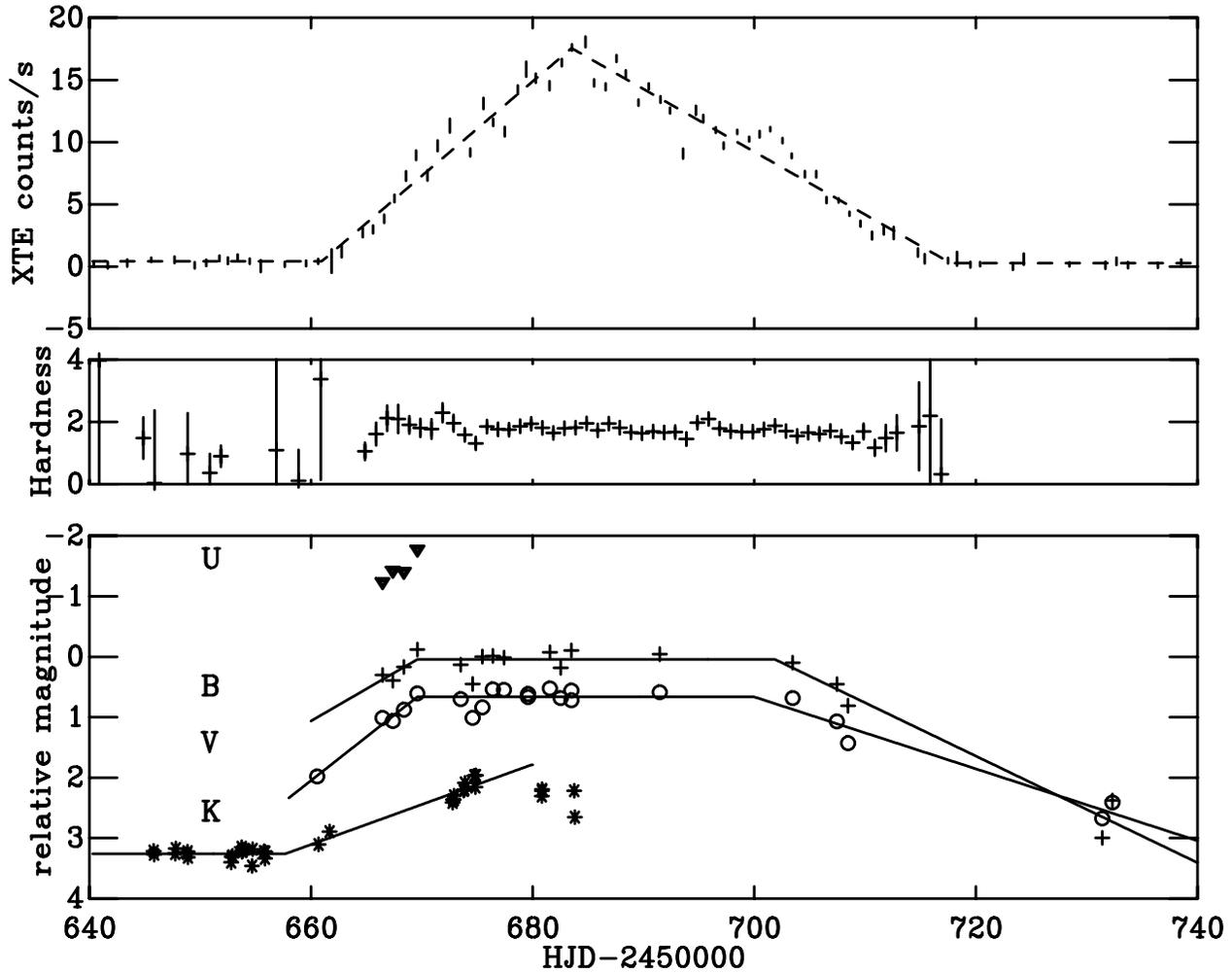}}
\caption{The top panel shows the $RXTE$ ASM X-ray light curve (2-12 keV)
for the August 1997 outburst of Aql X--1. Also shown is the simultaneous
fit to the linear rise and decay. The middle panel shows the ASM hardness
ratio (3.0--12.0keV/1.3--3.0keV).
The bottom panel shows the optical/IR magnitudes of Aql X--1 relative to a 
local standard (see section 2).}
\end{figure*}

\begin{figure*} 
\rotate[l]{\epsfxsize=500pt \epsfbox[00 00 700 750]{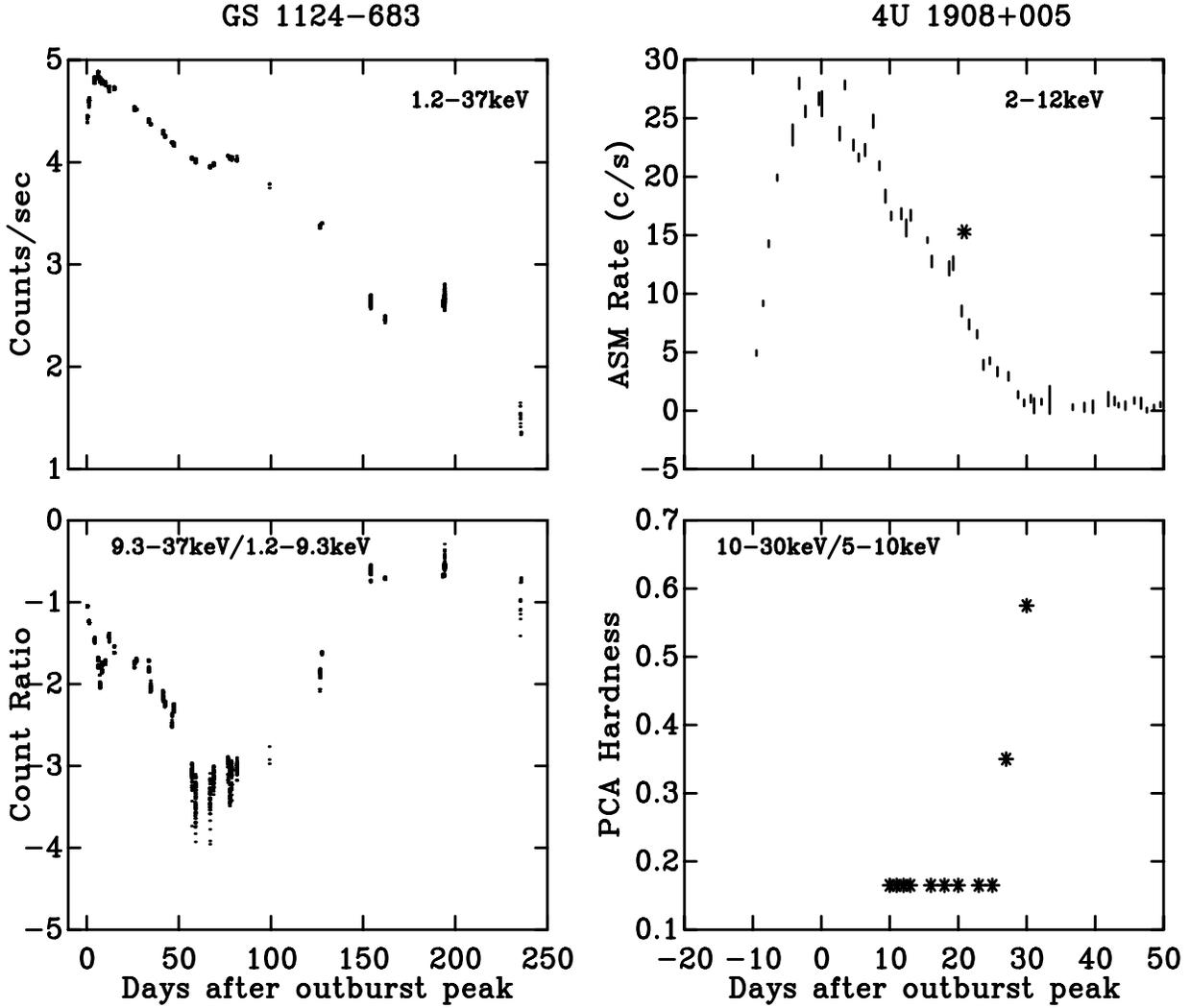}}
\caption{
Left figure: The spectral evolution of the January 1991 outburst of
GS~1124--68. The top and bottom panel show the LAC (1.2--3.7 keV) count
rate and the hardness (9.3--37 keV/1.2--9.3 keV) ratio. Right figure: The
spectral evolution of the February 1997 RXTE X-ray outburst of
Aql X--1. The top and bottom panel shows the RXTE/ASM (2--12 keV)
count rate and PCA hardness (10--30 keV/5--10 keV) ratio. The arrow shows
the time at which we expect to see a secondary maximum.  Both objects show a 
clear rise in the hardness ratio after the time of the secondary maximum. }
\end{figure*}

\end{document}